\documentclass[journal=jpccck,manuscript=article,layout=traditional]{achemso}
\usepackage{graphics}
\usepackage{color}
\usepackage[intlimits]{amsmath}
\usepackage{amssymb}
\usepackage{amsthm}
\usepackage{booktabs}
\usepackage{bm}

\author{R. Carmina Monreal}

\affiliation{Departamento de F\'{\i}sica Te\'{o}rica de la Materia Condensada and Condensed Matter Physics Center (IFIMAC), Universidad Aut\'{o}noma de Madrid, E-28049 Madrid, Spain}
\email{r.c.monreal@uam.es}

\title{Electron-Electron and Electron-Phonon Interactions in the Dynamics of Trap-Filling in Charged Quantum Dots}


\begin{document}

\begin{abstract}
We analyze theoretically the effects of electron-electron and electron-phonon interactions in the dynamics of a system of a few electrons that can be trapped to a localized state and detrapped to an extended band state of a small quantum dot (QD) using a simple model. In our model the QD is described by one or two single-particle energy levels while the trap is described by one single-particle level connected to the QD by a hopping Hamiltonian. Electron-electron Coulomb repulsion and electron-phonon interactions are included in the localized trap state.
In spite of its simplicity the time dependent model has no analytical solution but a numerically exact one can be found at a relatively low computational cost.
Using values of the parameters appropriate for defects in semiconductor QDs, we find that the electronic motion is quasi-periodic in time, with oscillations around mean values that are set in timescales of typically a few tenths of picoseconds to picoseconds.
We increase the number of electrons initially in the QD from one to four and find that one electron is transferred to the trap state for three electrons in the QD. At the more efficient values of the electron-phonon coupling, these characteristics are quite independent of the value of the electron-electron Coulomb repulsion in the trap, up to the value above which its infinite limit is reached. 
We conclude that strong electron-phonon interaction is an efficient mechanism that can provide the complete filling of a deep trap state on a sub to picoseconds timescale, faster than radiative exciton decay and Auger recombination processes. This leads to the complete suppression of the luminescence to the deep trap state and to the accumulation of electrons in the QD.

\end{abstract}

\maketitle

\newpage

\section{Introduction}

Optical properties of semiconductor nanocrystals (quantum dots QDs)\cite{Pichaardy, LEED3} are attracting an increasing interest for applications in optoelectronic devices\cite{Opto1,Opto2}, solar cells \cite{Solar1} and light emitting diodes \cite{LEED1,LEED2} among others.
 Although the core of the QD of a few nanometers in size has the perfect lattice symmetry of the bulk material, its band structure is discrete due to quantum-size effects. A consequence of quantum confinement of electrons is the phonon bottleneck: excited electrons are long lived excitations in a small QD because they would couple very inefficiently to phonons due to the large mismatch between the inter-level energy spacing and the optical phonon energy \cite{Nozik_PhysE, Yang_NatPhot}. 
Light emission in colloidal QDs is conditioned by the presence of defects and/or impurities at the surface. 
In these small systems, the large surface to volume ratio results in relative abundance of surface defects of varied microscopic nature, such as lattice relaxations, impurity atoms or surface ligands, generically called "traps", that produce states in the bandgap \cite{LEED3, JPCB_104_1715_vanDijken}.
Once the QD is excited by light across the bandgap, these localized states can trap charge carriers and alter the subsequent light emission. 
 A paradigmatic case is ZnO
 for which its luminescent properties are known to depend upon the atmosphere surrounding the particles
\cite{ JPCB_104_1715_vanDijken, ChemPhysLett_122_507_koch, JPhysChem_91_3789_bahnemann, JLumin_90_123_vanDijken, JPCB_104_4355_vanDijken}, 
 the charge on them \cite{JPCC_114_220_stroyuk, JPCC_115_21635_yamamoto,
JPCC_116_20633_cohn}, the passivation of surface defects \cite{JPCB_109_20810_norberg} or chemical modifications at the surface \cite{JACS_123_11651_shim}.
In ZnO two well separated emission bands appear: the narrow band (exciton band) is in the ultraviolet and 
originates from radiative recombination of excited electrons with the holes left in the valence band, while the usually prominent broad band 
is in the visible (green band) and arises 
from recombination of excited electrons with deep-trapped holes whose electronic states lie in the middle of the bandgap and are therefore localized. 
The relative intensity of both bands depends
on the above mentioned environmental conditions and, surprisingly, on the "charge" (ie. the excess of electrons in the conduction band)  on the nanocrystals \cite{JPCC_114_220_stroyuk, JPCC_115_21635_yamamoto,
JPCC_116_20633_cohn}. 
Photoexcitation of colloidal ZnO QDs under continuous illumination with ultraviolet radiation yields a quick quenching of the prominent green band 
while the excitonic band increases in intensity. Furthermore an infrared absorption band develops due to the accumulation of electrons in the conduction band,
that forms a ultralow-density electron gas \cite{JPCL_4_3024_faucheux, JPCL_5_976_faucheaux, ACSNano_8_1065_schimpf}. This is the problem we address in this paper.

A key point to understand the light emission properties of QDs is to know the timescales and efficiencies of possible mechanisms competing with radiative decay in the filling of the deep-trap state because if these mechanisms were much faster the photoluminescence will be quenched. 
One of the proposed mechanisms is a trap-assisted Auger recombination process in which a conduction band electron fills the deep-trap hole with the released energy transferred to another conduction band electron \cite{JPCC_116_20633_cohn, Nanolett_cohn}. Photoluminesce decay curves of ZnO yield times for this process that depend on crystal size, 
ranging from 60 ps to 300 ps for radius between 2 nm and 4 nm \cite{Nanolett_cohn}. 
These values are on the same order of magnitude as for other Auger recombination processes of electrons and holes in semiconductor nanocrystals. Auger times for trions in CdSe range between 5 ps and 1000 ps \cite{Nanolett_cohn_1, Nanolett_Vaxemburg}, 
while the times for Auger recombination of biexcitons  are from 5 ps to 500 ps for nanocrystals of 1 nm to 3 nm in size \cite{ChemRev_Guyot, Nanolett_Rabani, Science_Klimov, PRL_Htoon,PRB_Taguchi,Zhu}. 
Another mechanism has been explored in a recent calculation by du Foss\'e at al.  \cite{ChemMat_duFosse}, where the initial photoexcitation of an electron to the conduction band of a CdSe QD of ca. 2 nm in diameter leads to a transient trap state localized in a Cd-Cd dimer that appears and disappears on the picoseconds time scale. 
However, the existence of local lattice relaxations at the trap site strongly recalls electron-phonon interaction. Although the vibronic interaction has already been included in first-principles calculations of photoluminescence line shapes of a number of defects in bulk semiconductors  \cite{Janotti_PRL, Janotti_APL,Janotti_JAP}, all of these studies have in common that they focus on the static electronic or electronic plus vibronic configurations of the system. An equivalent analysis for a dynamical problem would be a formidable task nowadays even for a small system. Hence simple exactly solvable model Hamiltonians are useful to gain physical insight into the dynamics of electron-phonon interactions.


In a previous work \cite{PRB_Monreal} we analyzed the effects of electron-phonon interaction in the dynamics of a single electron 
that can be trapped to a localized state and detrapped to an extended band state, using a simple model. In the present paper we extend the analysis to a system of 2 to 4 electrons, focusing on the consequences of increasing the number of electrons in the QD. 
Upon photoexcitation, well characterized ZnO colloidal QDs in toluene,
 with sizes ranging from 2 to 6 nm in radii, can be charged all to the same maximum electron density of 
$\rho_e= (1.4 \pm 0.4) \times 10^{20}$ cm$^{-3}$ \cite{ACSNano_8_1065_schimpf}. Hence, at the lower limit of this density, nanospheres of 2 and 3 nm in radii will have a maximum of 3 and 11 electrons, respectively. 
In our model the QD is described by one or two single-particle energy levels.  
Substitution of a real QD by a few discrete energy levels has been frequent in the literature, in particular in the analysis of the phonon bottleneck problem
\cite{Inoshita_PRB, Kral_PRB,Stauber_PRB, Vasilevskiy_PRB}. This description will be valid in situations where a few levels could be populated with appreciable probability in the initial photoexcitation event (as for example for photon energies slightly above the bandgap) or by electronic hopping from the trap. 
The trap is modeled by one single-particle energy level well below the conduction band,
 as is the case for deep traps in semiconductor QDs
\cite{JPCB_104_1715_vanDijken, JLumin_90_123_vanDijken, JPCB_104_4355_vanDijken, JPCB_109_20810_norberg,Janotti_PRL,Janotti_APL}, 
and the QD and the trap are connected by a hopping Hamiltonian.
In the photoexcitation of a QD electrons are excited to the conduction band. We take this event as our starting point and consider a number of electrons in our energy levels at time t=0. The electrons are subsequently allowed to hop back and forth to the trap and we calculate numerically the evolution of the coupled electron-phonon system as a function of time. The physical parameters of our problem are in principle unrestricted but we will use values appropriate for defects in semiconductors QDs, taken from Refs. \cite{Janotti_PRL,Janotti_APL,Janotti_JAP}.
This minimal model allows us to investigate numerically both the short and the long time dynamics of the electrons, which will be presently infeasible for first-principles type of Hamiltonians.

Natural units $\hbar=e=m_e=k_{B}=1$ are used except otherwise indicated.

\section{Theoretical Methods}
\label{sec-theory}

We use a type of Anderson-Holstein impurity model  in which the quantum dot is described by one single-particle level of energy $\epsilon_0$ or two single-particle levels of energies 
$\epsilon_0$ and $\epsilon_1$ ($\epsilon_1 > \epsilon_0$), while the trap
has one single-particle level of energy $\epsilon_T$ well below $\epsilon_0$. 
Electron-electron Coulomb repulsion $U$ between two electrons of opposite spins is considered in the trap localized state but it is neglected in the dot 
since electrons move in extended levels in real QDs. 
The interaction between an electron in the dot and another in the trap should be smaller than $U$ and it is also neglected.
  Although several phonons could be active at the trap site, first-principles calculations of the photoluminescence lineshapes of several defects in semiconductors \cite{Janotti_PRL,Janotti_APL,Janotti_JAP}, 
show that two phonons of similar energies and coupling constants are enough to fit the experiments. Then, in our model electrons in the trap couple to a single local phonon of energy
 $\omega_0$ with a coupling constant $\lambda$ \cite{Holstein}.
Electron-phonon interaction is neglected in the QD because of the bottleneck effect: excited electrons are long lived excitations in a small QD as
 they would couple very inefficiently to phonons due to the large mismatch between the inter-level energy spacing and the phonon energy. QD and trap are connected via a hopping parameter $V_{DT}$.
In the more general case of a two-level QD connected to the trap, the Hamiltonian is written as 

\begin{equation}
\hat H=\hat H_D+ \hat H_T+\hat H_{hop},
\label{Ham}
\end{equation}
where $\hat H_D$ and $\hat H_T$  are the Hamiltonians of the uncoupled dot and trap subsystems respectively 

\begin{equation}
\hat H_{D}=\epsilon_{0} \sum_{\sigma}\hat n_{0 \sigma}+\epsilon_{1} \sum_{\sigma}\hat n_{1 \sigma},
\label{Ham-D}
\end{equation}

\begin{equation}
\hat H_{T}= \epsilon_T \sum_{\sigma}\hat n_{T \sigma}+ U \hat n_{T \uparrow} \hat n_{T \downarrow}+
\omega_0 \hat b^{\dagger} \hat b+ \lambda \sum_{\sigma} \hat n_{T \sigma}(\hat b^{\dagger}+\hat b),
\label{Ham-T}
\end{equation}

and the hopping Hamiltonian $\hat H_{hop}$ is written as
\begin{equation}
\hat H_{hop}=V_{DT} \sum_{\sigma} (\hat c_{T \sigma}^{\dagger} \hat c_{0, \sigma} +\hat c_{T \sigma}^{\dagger} \hat c_{1 \sigma} +hc.).
\label{Ham-hop}
\end{equation}

\begin{figure}
\centering
\includegraphics[width=100mm]{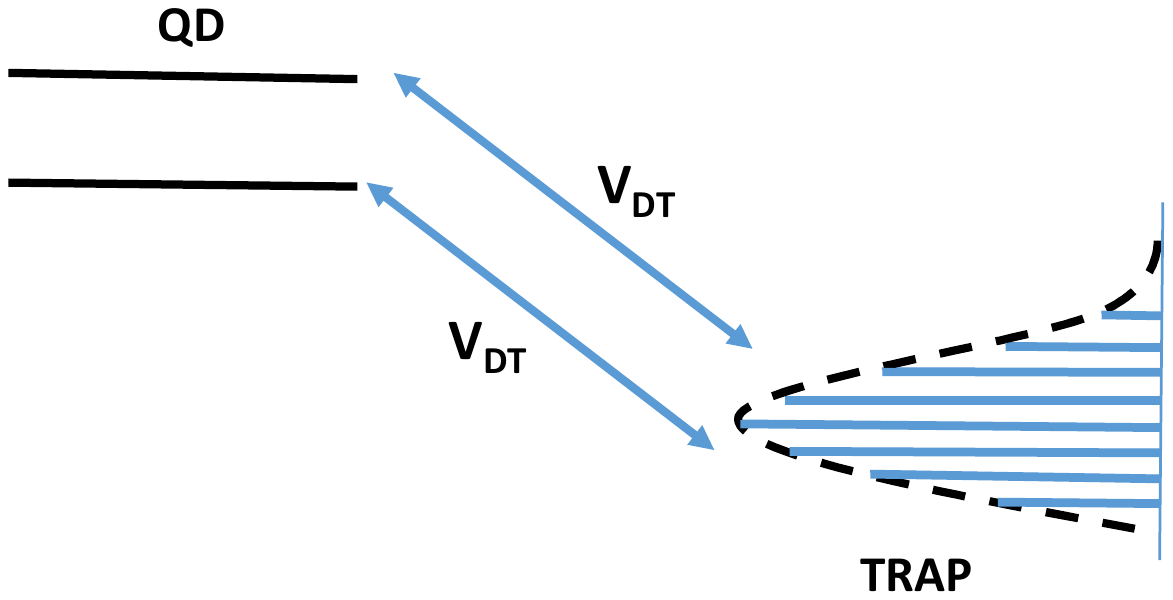}
\caption{ The scheme of our model Hamiltonian. The QD is modeled by  two single-particle energy levels while the trap, coupled to a local phonon, presents an infinite number of sublevels equally spaced by the phonon energy, with weights distributed according to a Poisson distribution function. Electrons in the QD can jump back and forth to the trap via a hopping parameter $V_{DT}$
.} 
\label{esq1}
\end{figure}

In eqs \ref{Ham-D}, \ref{Ham-T} and \ref{Ham-hop}  $\hat c_{0 \sigma}$ ($\hat c_{0 \sigma}^{\dagger}$), $\hat c_{1 \sigma}$ ($\hat c_{1 \sigma}^{\dagger}$) and 
$\hat c_{T}$ ($\hat c_{T}^{\dagger}$)
are the annihilation (creation) operators for electrons of spin $\sigma$ in each of the single-particle levels, 
$\hat n_{0 \sigma}=\hat c_{0 \sigma}^{\dagger} \hat c_{0 \sigma}$, $\hat n_{1 \sigma}=\hat c_{1 \sigma}^{\dagger} \hat c_{1 \sigma}$ and 
$\hat n_{T \sigma}=\hat c_{T \sigma}^{\dagger} \hat c_{T \sigma}$ being their respective number operators,
$\hat b$ ($\hat b^{\dagger}$) are the annihilation (creator) operators for phonons, $\hat b^{\dagger} \hat b$ being their number operator.
 The hopping parameter in eq \ref{Ham-hop}  has the same value for electrons in $\epsilon_{0}$ and $\epsilon_{1}$, for simplicity.
Obviously, the Hamiltonian for a single-level QD is obtained by neglecting level $\epsilon_1$ in eqs \ref{Ham-D} and \ref{Ham-hop}.
Our model system is schematically drawn in Figure \ref{esq1}. 

For a better understanding of our results we quote here the so called atomic limit ($V_{DT} \rightarrow 0$) in which an electron in a localized level interact with phonons.
The stationary solution of eq \ref{Ham-T} \cite{Mahan} yields to renormalization of the electron energy to $\tilde \epsilon_T=\epsilon_T-\frac{\lambda^2}{\omega_0}$ 
and of the Coulomb repulsion $U$ to $U_{eff}=U-2\frac{\lambda^2}{\omega_0}$
and a local density of states which, at zero temperature, reads 

\begin{eqnarray}
\rho_{T \sigma}(\omega)=-\frac{1}{\pi} \lim_ {\eta \rightarrow 0} Im e^{-g} \sum_{n=0}^{\infty}  \frac{g^n}{n!} 
[ \frac{\langle(1-\hat n_{T \sigma})(1-\hat n_{T -\sigma}) \rangle}{\omega-\tilde \epsilon_T-n \omega_0+i\eta}
+\frac{\langle(1-\hat n_{T -\sigma})\hat n_{T \sigma} \rangle}{\omega-\tilde \epsilon_T+n \omega_0+i\eta} \nonumber \\
+\frac{\langle(1-\hat n_{T \sigma}) \hat n_{T -\sigma} \rangle}{\omega-\tilde \epsilon_T-U_{eff}-n \omega_0+i \eta}
+\frac{\langle\hat n_{T \sigma} \hat n_{T -\sigma} \rangle}{\omega-\tilde \epsilon_T-U_{eff}+n \omega_0+i\eta} ],
\label{rho}
\end{eqnarray}
where $g=(\frac{\lambda}{\omega_0})^2$ is the Huang-Rhys parameter and $Im$ stands for imaginary part.
The density of states consists of peaks (sublevels) at
$\tilde \epsilon_T  \pm n \omega_0$ and  $\tilde \epsilon_T+U_{eff} \pm n \omega_0$ with weights distributed according to the Poisson distribution function. 
This structure will be seen in our results. The effect of temperature is to transfer spectral weight from the sublevels with $n \simeq g$, where the density of states has a maximum at $T=0$, to the sublevels with small and large values of $n$  \cite{Mahan} . The effect depends exponentially on the ratio $\frac{\omega_0}{T} $ and is small  for  $\frac{\omega_0}{T} \ge 1$, which is the case at room temperature and below for the value of $\omega_0$ we will use in the calculations. Also, thermal excitation within the QD levels is small 
in our case because  $\frac{\epsilon_1-\epsilon_0}{T} \ge 1$.
 For these reasons and for the sake of simplicity, we will work a zero temperature.
We note that, while $U$ is a positive defined energy, $U_{eff}$ can be negative indicating an effective attractive 
interaction between electrons mediated by the electron-phonon interaction.

\begin{figure}
\centering
\includegraphics[width=100mm]{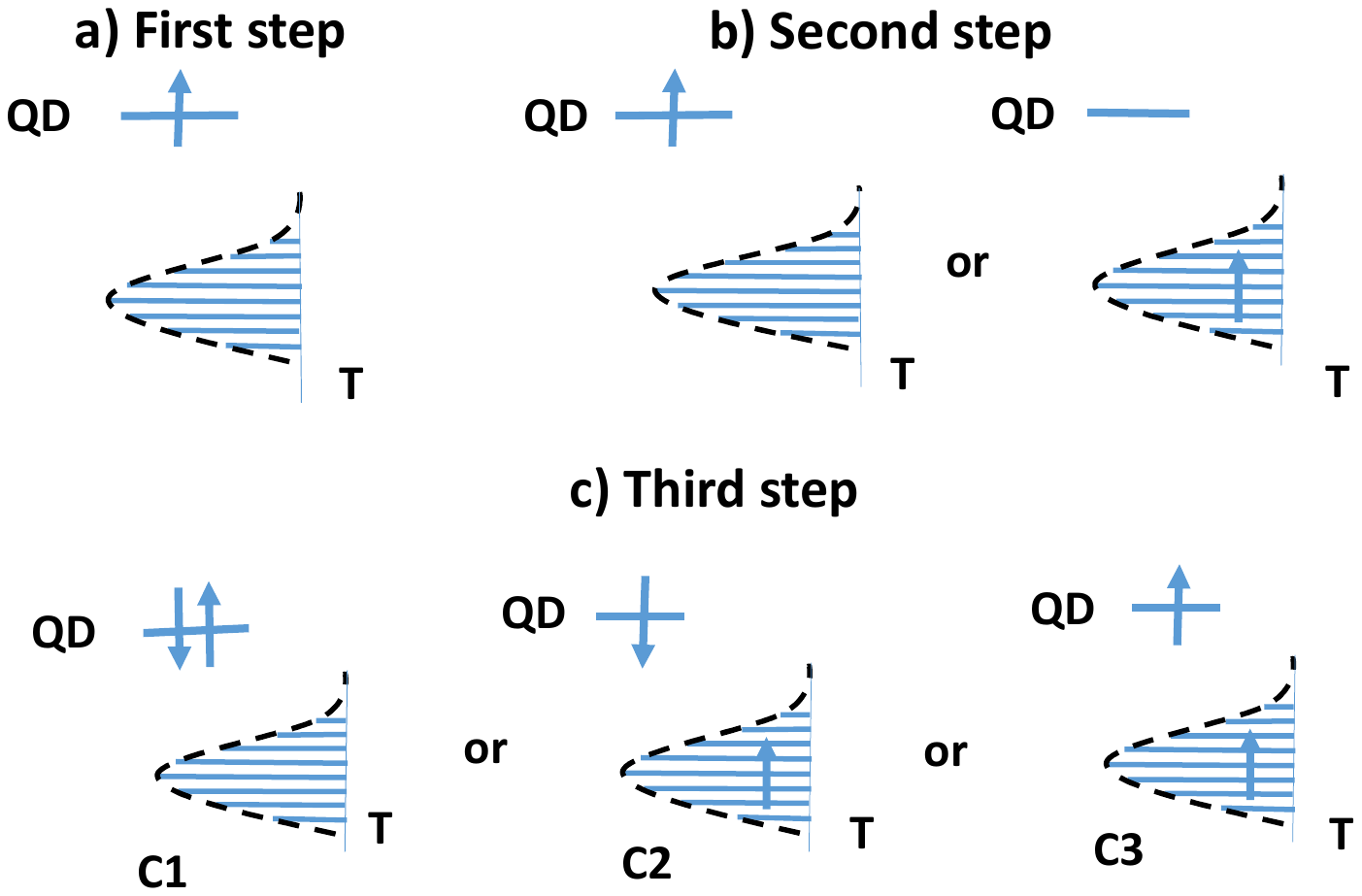}
\caption{ The scheme for a three-step process in a system of two electrons in a single-level QD connected to the trap T.  In the first step (a) a spin up electron is put in the QD while the trap state is empty. In the second step (b) this electron can be found in the trap (b) right) with probability $p_1$ or in the dot (b) left) with probability $1-p_1$. In the third step (c) another electron, with up or down spin, is put in the dot, yielding the three possible initial configurations drawn in the Figure. Configuration $C1$ identical to the one-step process.
} 
\label{esq2}
\end{figure}

To solve the time dependent problem for Hamiltonian eq \ref{Ham} for any value of the parameters, we use a complete basis set containing all possible states of electrons and phonons. 
The time dependent wave function of the system is written as a linear combination of these states and the time-dependent Sch\"ordinger equation is projected onto each of them
leading to a system of coupled linear differential equations for the combination coefficients that is solved numerically with given initial conditions. 
We will consider two possible
scenarios for the photoexcitation process that will be translated to the initial conditions. In a one-step process two (or more) electrons are put in the QD at time $t=0$ (configuration $C1$ in Figure \ref{esq2} for two electrons)
and we let the system to evolve thereafter. A three-step process is schematized in Figure \ref{esq2}. In the first step (a) one electron is put in the dot. 
Our previous work \cite{PRB_Monreal} tells us that the system evolves,
in a subpicoseconds timescale, 
towards a quasi-stationary regime, that we call second  step (b), in which the electron has a probability $p_1$ of being transferred to the trap with $n_{b1}$ phonons around
( b) right) and a probability $(1-p_1)$ of remaining in the first step configuration ( b) left). 
In the third step (c) a second electron, with up or down spin, is put in the dot at time $t=0$ that can find the first one in the dot or in the trap, yielding configurations $C1$, $C2$ or $C3$.
 Configuration $C1$  is identical to the one-step process. Thus, 
while for the one-step process we only need one calculation for the initial configuration $C1$, for the three-step process we need to consider in addition the initial configurations schematized in $C2$ and $C3$ and combine the results with weights $1-p_1$ and $p_1$ as appropriate. 

Consideration of the three-step process is motivated by the fact that the 
timescale for electron-electron and electron-phonon interactions is much shorter than that of radiative processes and, moreover, it would be the more probable process under continuous illumination with low power lasers.
We will first analyze the simplest case of two electrons in a single-level QD followed by the analysis of two electrons in the double-level QD, where the electrons can visit more single-particle states, considering one-step and three-steps processes in both cases. We will  show  that the results of both calculations 
do not differ substantially and for this reason only the one-step model will be considered in systems with three and four electrons. Note that the single-level QD only admits two electrons and, consequently, it is essential to consider more than one single particle level if we want to study the effect that an increasing number of electrons in the QD has in the trap occupancy.

\subsection{A single-level QD connected to the trap: theory} 
\label{sec-2sites}


\begin{figure}
\centering
\includegraphics[width=100mm]{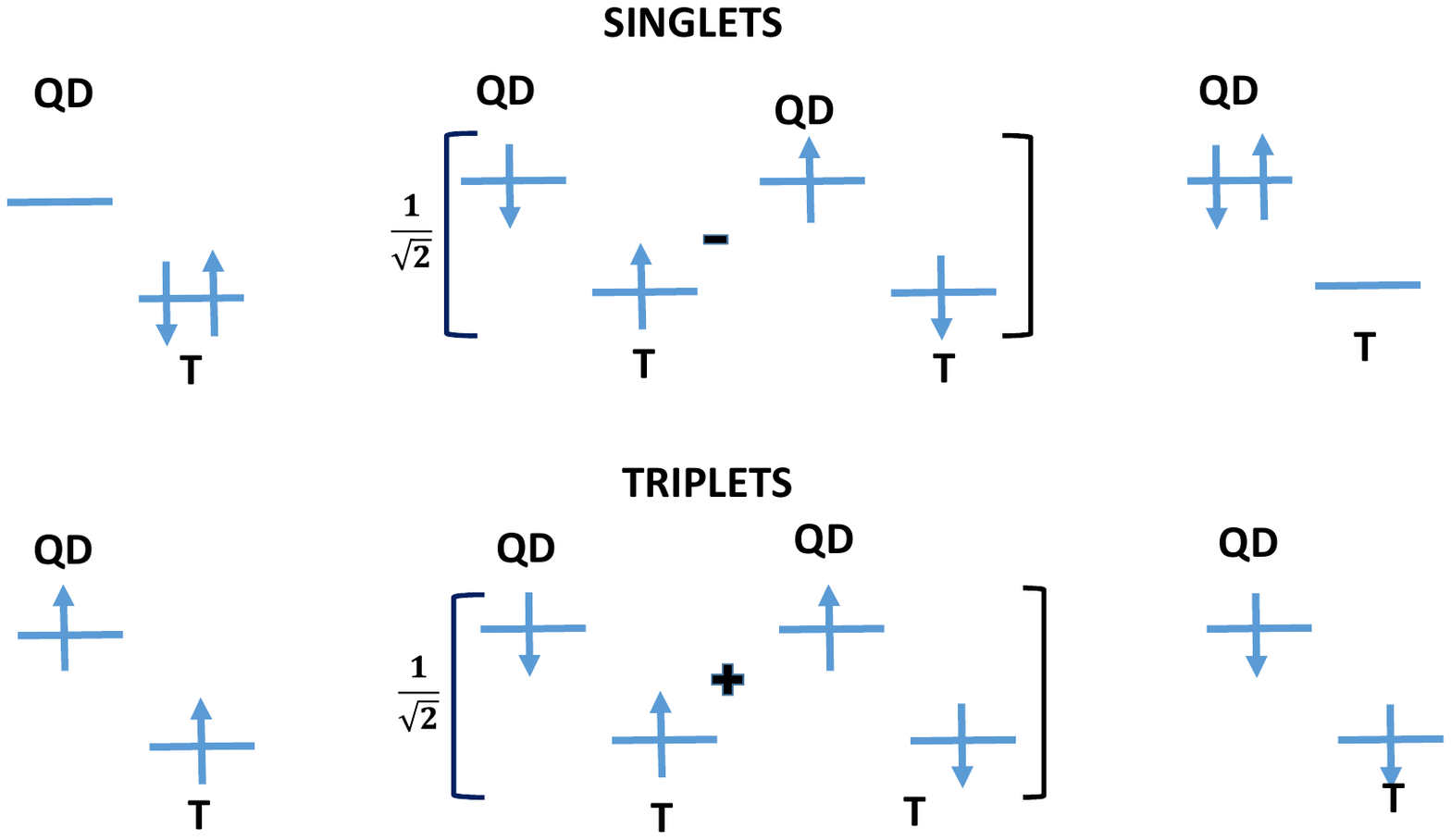}
\caption{The scheme of the possible configurations of two electrons in a single-level QD connected to the trap T. The multiple levels associated to the trap are not drawn, for clarity. The upper row schematizes the singlets with two electrons in the trap (left), one electron in the dot and the other in the trap in the singlet combination (middle) and two electrons in the dot (right). The lower row schematizes the triplets, all with one electron in the dot and the other in the trap, with $S_z=1$ (left), $S_z=0$ (middle) and $S_z=-1$ (right).
} 
\label{esq3}
\end{figure}

We start considering the simplest system consisting of two electrons in a single-level QD connected to the trap. 
Note that the Hamiltonian of eq. 1 commutes with  $S^2$ and  $S_z$, $S$ being the total spin and $S_z$ its z-component. Hence $S$ and $S_z$ are conserved, and the singlet and the triplet states of the two-electron system do not mix.
The states of this system are described using a basis set written in the number representation as
\begin{equation} 
|n_{0 \uparrow} n_{0 \downarrow};  n_{T \uparrow} n_{T \downarrow}\rangle \otimes |n \rangle,
\label{n}
\end{equation}
with $n_{0 \sigma}$ and $n_{T \sigma}$ being the occupation numbers of electrons of spin $\sigma$ in the single-particle levels $\epsilon_0$
and $\epsilon_T$, respectively and n being the phononic number. In this representation the three singlet states  are

\begin{eqnarray}
|\varphi_{TT,n}\rangle &= &|0 0; 1 1\rangle \otimes|n \rangle , \nonumber \\ 
|\varphi_{DT,n}\rangle &= &\frac{1}{\sqrt 2}(|1 0; 0 1\rangle - |0 1; 1 0\rangle) \otimes|n \rangle,  \nonumber \\ 
|\varphi_{DD,n}\rangle &= &|1 1; 0 0\rangle \otimes|n \rangle,  \nonumber \\
\label{basis-2}
\end{eqnarray}
representing two electrons in the trap, one electron in the dot and the other in the trap and two electrons in the dot, that are schematically drawn in the upper row of Figure \ref{esq3} from left to right, respectively. These singlet states are coupled by $H_{hop}$ and, consequently, evolve in time. The three triplet states, schematically drawn in the lower row of Figure \ref{esq3}, are uncoupled themselves and uncoupled to the singlets so they do not evolve in time.

Thus, the time-dependent wave function is written as a linear combination of the singlet states

\begin{equation}
|\psi(t) \rangle= \sum_{n=0}^{\infty}[a_{DD,n}(t)|\varphi_{DD,n}\rangle +a_{TT,n}(t)|\varphi_{TT,n}\rangle +a_{DT,n}(t)|\varphi_{DT,n}\rangle],
\label{psi2-t}
\end{equation}
and the time-dependent Sch\"ordinger equation for Hamiltonian eq \ref{Ham} is projected onto the basis, leading to the following system of coupled linear differential equations for the combination coefficients

\begin{eqnarray}
\frac{d a_{DD,n}(t)}{dt}&=&-i(2\epsilon_0+n\omega_0)a_{DD,n}(t)-i \sqrt 2 V_{DT} a_{DT,n}(t), \nonumber \\
\frac{d a_{TT,n}(t)}{dt}&=&-i(2 \epsilon_T+U+n\omega_0)a_{TT,n}(t)-i \sqrt 2 V_{DT} a_{DT,n}(t) \nonumber \\
                        & &-i 2 \lambda \sqrt{n}a_{TT,n-1}(t)-i 2 \lambda \sqrt{n+1}a_{TT,n+1}(t), \nonumber \\ 
\frac{d a_{DT,n}(t)}{dt}&=&-i(\epsilon_0+\epsilon_T+n\omega_0)a_{DT,n}(t)-i \sqrt 2 V_{DT} [a_{DD,n}(t)+ a_{TT,n}(t)] \nonumber \\
                        & &-i \lambda \sqrt{n}a_{DT,n-1}(t)-i \lambda \sqrt{n+1}a_{DT,n+1}(t).
\nonumber \\
\label{equ2-t}
\end{eqnarray}

For the one-step process, where two electrons (of opposite spins) start in the dot at $t=0$ (see configuration C1 in Figure \ref{esq2}), 
the initial conditions are 
$a_{DD,n}(t=0)=\delta_{n,0}$, $a_{TT,n}(t=0)=0$ and $a_{DT,n}(t=0)=0 \; \forall n$. 
The solution of  eq \ref{equ2-t} give us the time-dependent occupancies of the trap and the dot and the number of phonons as
\begin{equation}
n_{T}(t)\equiv \langle \psi(t)|\sum_{\sigma}\hat{n}_{T \sigma}|\psi(t) \rangle =\sum_{n=0}^{\infty}(2| a_{TT,n}(t)|^2+|a_{DT,n}(t)|^2),
\label{nT2}
\end{equation}

\begin{equation}
n_{D}(t)\equiv \langle \psi(t)|\sum_{\sigma}\hat{n}_{0 \sigma}|\psi(t) \rangle =\sum_{n=0}^{\infty}(2 | a_{DD,n}(t)|^2+|a_{DT,n}(t)|^2),
\label{nD}
\end{equation}
and 
\begin{equation}
n_{b}(t)\equiv \langle \psi(t)|\hat{b}^{\dagger} \hat{b}|\psi(t) \rangle =\sum_{n=0}^{\infty} n(|a_{TT,n}(t)|^2+|a_{DD,n}(t)|^2+|a_{DT,n}(t)|^2),
\label{nb}
\end{equation}
respectively. The correct normalization of the wavefunction of eq \ref{psi2-t} guarantees conservation of the number of electrons:  $n_{T}(t)+n_{D}(t)=2 \; \forall t$. 

It is also interesting to look at the probabilities $p_{TT}(t), p_{DD}(t)$, and $p_{DT}(t)$ of
finding two electrons in the trap, two electrons in the dot and one electron in the dot and the other in the trap at time $t$.
These are defined as
\begin{equation}
p_{TT}(t)=\sum_{n=0}^{\infty}| a_{TT,n}(t)|^2,
\end{equation}

\begin{equation}
p_{DD}(t)=\sum_{n=0}^{\infty}| a_{DD,n}(t)|^2,
\end{equation}
and

\begin{equation}
p_{DT}(t)=\sum_{n=0}^{\infty}| a_{DT,n}(t)|^2,
\end{equation}
respectively, with $n_T(t)=2p_{TT}(t)+p_{DT}(t)$ and  $n_D(t)=2p_{DD}(t)+p_{DT}(t)$, as it should.

While for the one-step process these magnitudes are obtained directly from one calculation as stated above, 
the three-step process requires to consider the possible different configurations in the third step depicted in Figure \ref{esq2}. Configuration $C1$ is identical to the one-step process and occurs with probability $1-p_1$. The two configurations occurring with probability $p_1$ are $C2$ and $C3$. 
Configuration $C2$ is the superposition of the singlet $|\varphi_{DT}\rangle$ and the triplet state with $S_z=0$, both with weights ${1 \over \sqrt2}$. 
 Then, since the triplet states do not evolve in time, we still need to run another calculation starting with the singlet  $|\varphi_{DT}\rangle$
 with  initial conditions
$a_{DT,n}(t=0)=\delta_{n,n_{b1}}$, $a_{DD,n}(t=0)=0$ and $a_{TT,n}(t=0)=0 \; \forall n$, yielding an occupancy  $n_{Ti}(t)$, so the trap occupancy for the initial configuration $C2$ 
is $\frac{n_{Ti}(t)+1}{2}$. Configuration $C_3$ is the triplet with $S_z=1$ and do not evolve in time.
Finally,  the trap occupancy for the three-step process 

\begin{equation}
n_{T3}(t)=(1-p_{1})n_{T1}(t)+p_{1} \frac{\frac{1}{2}n_{Ti}(t)+\frac{3}{2}}{2},
\end{equation} 
where $n_{T1}(t)$ is the occupancy calculated for the one-step process.

We will also calculate time-averaged values of occupations and probabilities because these are the physically 
relevant values for other slower processes competing with hopping, electron-electron and electron-phonon interactions in the filling of the trap, such as light emission.
As an example, the time-averaged trap occupancy is calculated as
\begin{equation}
\langle n_{T} \rangle =\frac{1}{t_{max}} \int_{0}^{t_{max}} dt\; n_T(t),
\label{nT-taver}
\end{equation}
where $t_{max}$ is the maximum value of time in our calculations (usually $\omega_{0}t_{max}=200$). 
It will be apparent from the results that we will present below that much smaller values of $t_{max}$ could be used to give an accurate value of $\langle n_{T} \rangle$. 
Time averaging of the probabilities is performed in the same way.

\subsection{ A two-level QD connected to the trap: theory}

We now consider a QD having two single-particle levels of energies $\epsilon_0$ and $\epsilon_1$ ($\epsilon_1 > \epsilon_0$). 
To solve the time-dependent problem our basis of states will be written in the occupation number representation as

\begin{equation}
|n_{0 \uparrow} n_{0 \downarrow}; n_{1 \uparrow} n_{1 \downarrow}; n_{T \uparrow} n_{T \downarrow}\rangle \otimes |n \rangle,
\label{basis-3}
\end{equation} 
with $n_{0 \sigma}$, $n_{1 \sigma}$ and $n_{T \sigma}$ being the occupation numbers of electrons of spin $\sigma$ in the single-particle levels $\epsilon_0$, 
$\epsilon_1$ and $\epsilon_T$, respectively and $n$ being the phononic number. Since the possible states in the basis depend on the number of electrons we put in the system we detail first the most involved case of having two electrons. 

\begin{figure}
\centering
\includegraphics[width=100mm]{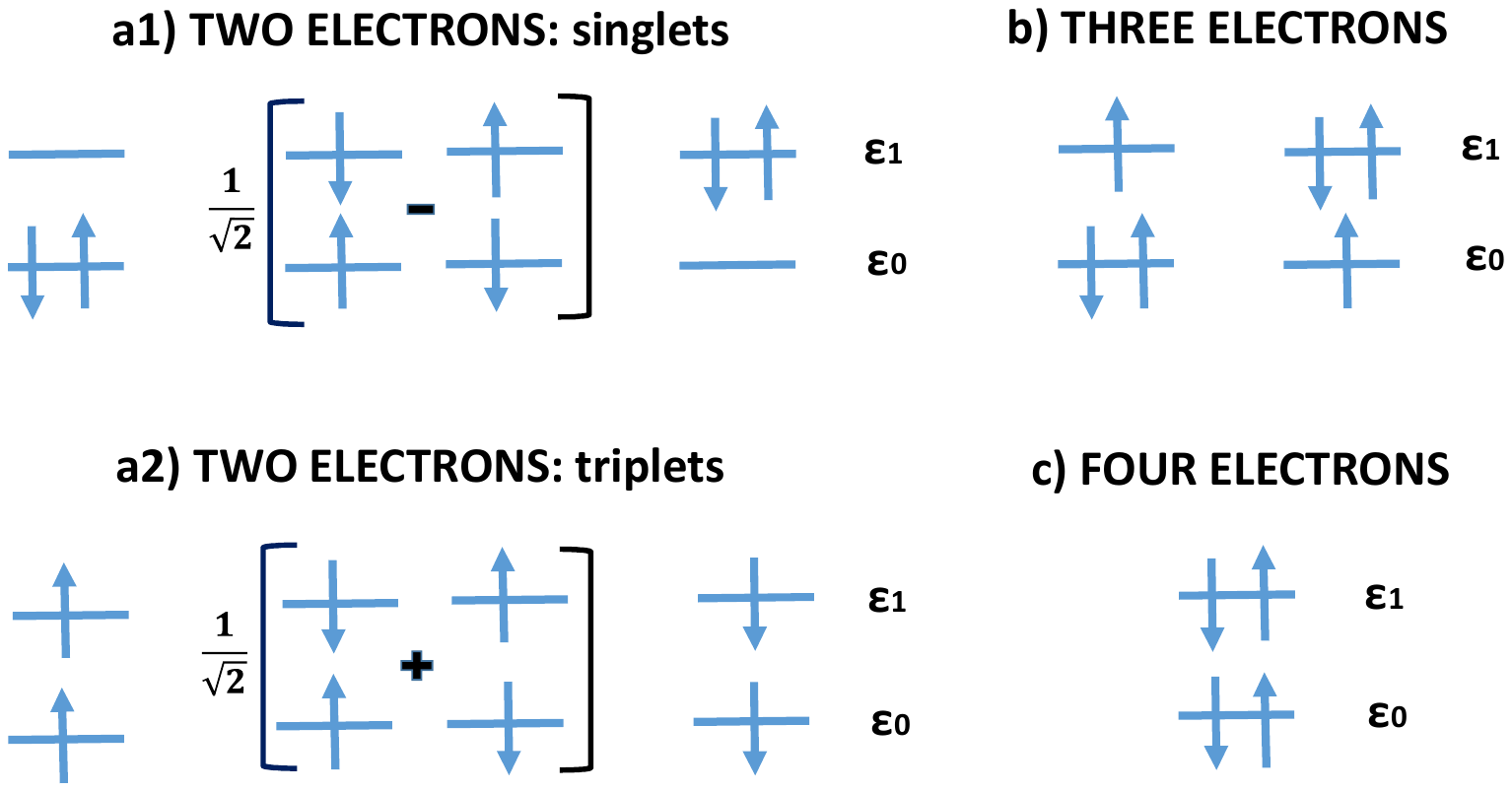}
\caption{ The possible configurations of two (a), three (b) and four (c) electrons in a two-level QD with single-particle energy levels $\epsilon_0$ and $\epsilon_1$
.} 
\label{esq4}
\end{figure}

Similar to the case of the single-level QD, the full problem can be separated in two independent ones, one for the singlet and the other for the triplet states. 
The singlet states of the two-electron system that are coupled by $H_{hop}$ and, therefore evolve in time, are

\begin{eqnarray}
|\varphi_{DD0,n}\rangle &= &|1 1; 0 0; 0 0\rangle \otimes|n \rangle,  \nonumber \\
|\varphi_{DD1,n}\rangle &= &\frac{1}{\sqrt 2}(|1 0; 0 1; 0 0\rangle - |0 1; 1 0; 0 0\rangle) \otimes|n \rangle,  \nonumber \\ 
|\varphi_{DD2,n}\rangle &= &|0 0; 1 1; 0 0\rangle \otimes|n \rangle,  \nonumber \\
|\varphi_{TT,n}\rangle &= &|0 0; 0 0; 1 1\rangle \otimes|n \rangle,  \nonumber \\ 
|\varphi_{D0T,n}\rangle_S &= &\frac{1}{\sqrt 2}(|1 0; 0 0; 0 1\rangle - |0 1; 0 0 ;1 0\rangle) \otimes|n \rangle,  \nonumber \\
|\varphi_{D1T,n}\rangle_S &= &\frac{1}{\sqrt 2}(|0 0; 1 0; 0 1\rangle - |0 0; 0 1; 1 0\rangle) \otimes|n \rangle.  \nonumber \\  
\label{Singlets}
\end{eqnarray}

The first three lines of eq \ref{Singlets}, with two electrons in the dot, are schematized in part $a1$) of Figure \ref{esq4} from left to right, respectively. The fourth line represents the state with the two electrons in the trap and the last two lines represent the two singlet states obtained by hopping of any of the two electrons from dot to trap or viceversa.
The time-dependent wave function is then written as a linear combination of the basis states

\begin{eqnarray}
|\psi_{S}(t) \rangle = \sum_{n=0}^{\infty}&[&a_{DD0,n}(t)|\varphi_{DD0,n}\rangle +a_{DD1,n}(t)|\varphi_{DD1,n}\rangle +a_{DD2,n}(t)|\varphi_{DD2,n}\rangle \nonumber \\
                                         & &+a_{TT,n}(t)|\varphi_{TT,n}\rangle +a_{D0T,n}(t)|\varphi_{D0T,n}\rangle_S+a_{D1T,n}(t)|\varphi_{D1T,n}\rangle_S],
\label{psiSinglets-t}
\end{eqnarray}
and the time-dependent Sch\"ordinger equation for Hamiltonian eq \ref{Ham} is projected onto each of the basis states, leading to the system of coupled linear differential equations given by eq S1 in the Supporting Information.


The triplet states with $S_{z}=0$ that are coupled by $H_{hop}$ are

\begin{eqnarray}
|\varphi_{DD,n}\rangle_{T0} &= &\frac{1}{\sqrt 2}(|1 0; 0 1; 0 0\rangle + |0 1; 1 0; 0 0\rangle) \otimes|n \rangle,  \nonumber \\ 
|\varphi_{D0T,n}\rangle_{T0} &= &\frac{1}{\sqrt 2}(|1 0; 0 0; 0 1\rangle + |0 1; 0 0 ;1 0\rangle) \otimes|n \rangle,  \nonumber \\
|\varphi_{D1T,n}\rangle_{T0} &= &\frac{1}{\sqrt 2}(|0 0; 1 0; 0 1\rangle + |0 0; 0 1 ;1 0\rangle) \otimes|n \rangle.  \nonumber \\
\label{Triplets0}
\end{eqnarray}
The state in the first line of eq  \ref {Triplets0} is depicted in Figure \ref{esq4}, a2) middle and the other two states are obtained from this one by hopping of any of the two electrons to the trap.

The triplets with $S_z=+1$ coupled by $H_{hop}$  are

\begin{eqnarray}
|\varphi_{DD,n}\rangle_{T1} &= &|1 0; 1 0; 0 0\rangle \otimes|n \rangle,  \nonumber \\
|\varphi_{D0T,n}\rangle_{T1} &= &|1 0; 0 0; 1 0\rangle \otimes|n \rangle,  \nonumber \\
|\varphi_{D1T,n}\rangle_{T1} &= & |0 0; 1 0; 1 0\rangle \otimes|n \rangle,  \nonumber \\
\label{Tripletequ2-t}
\end{eqnarray}
and the corresponding basis for $S_{z}=-1$.  The first line in eq  \ref{Tripletequ2-t} is depicted in Figure \ref{esq4}, a2)  left and the other two states are obtained from this one by hopping of any of the two electrons to the trap. 
Since the matrix elements of $H_{hop}$ within each subspace of given $S_z$ are identical, 
the wave function within the subspace 
is then written as the corresponding linear combination of the basis states

\begin{eqnarray}
|\psi_{Tz}(t) \rangle = \sum_{n=0}^{\infty}[b_{DD,n}(t)|\varphi_{DD,n}\rangle_{Tz} +b_{D0T,n}(t)|\varphi_{D0T,n}\rangle_{Tz}+b_{D1T,n}(t)|\varphi_{D1T,n}\rangle_{Tz}],
\label{psiTriplets-t}
\end{eqnarray}
with the coefficients in the linear combination, that do not depend on $S_z$, fulfilling the set of coupled differential equations given by eq S2 in the Supporting Information.


For the one-step process, where the two electrons start in the dot at $t=0$, we have three possible singlet initial states, 
namely $|S_0\rangle \equiv |\varphi_{DD0,n=0}\rangle$, $|S_1\rangle \equiv |\varphi_{DD1,n=0}\rangle$ or 
$|S_2\rangle \equiv|\varphi_{DD2,n=0}\rangle$, with energies $2\epsilon_0$, $\epsilon_0+\epsilon_1$ and $2 \epsilon_1$ respectively, and 
the three triplet states $|\varphi_{DD,n=0}\rangle_{T_0}$, $|\varphi_{DD,n=0}\rangle_{T_1}$
or $|\varphi_{DD,n=0}\rangle_{T_{-1}}$ all with energy $\epsilon_0+\epsilon_1$.
For each of the initial singlets, numerated by $i=0, 1, 2$, we solve the set of eq S1 with the corresponding initial condition, yielding 
the probabilities of finding both electrons in the trap, $p_{TT}^{(i)}(t)$, 
both electrons in the dot, $p_{DD}^{(i)}(t)$, one electron in the dot and the other in the trap, $p_{DT}^{(i)}(t)$ and trap occupancy, $n_{T}^{(i)}(t)$, as

\begin{equation}
p_{TT}^{(i)}(t)=\sum_{n=0}^{\infty}| a_{TT,n}^{(i)}(t)|^2,
\end{equation}

\begin{equation}
p_{DD}^{(i)}(t)=\sum_{n=0}^{\infty}| a_{DD,n}^{(i)}(t)|^2,
\end{equation}

\begin{equation}
p_{DT}^{(i)}(t)=\sum_{n=0}^{\infty}[| a_{D0T,n}^{(i)}(t)|^2+| a_{D1T,n}^{(i)}(t)|^2].
\end{equation}
and

\begin{equation}
n_{T}^{(i)}(t)\equiv \langle \psi_{S_i}(t)|\sum_{\sigma}\hat{n}_{T \sigma}|\psi_{S_i}(t) \rangle =\sum_{n=0}^{\infty}
[2| a_{TT,n}^{(i)}(t)|^2+|a_{D0T,n}^{(i)}(t)|^2+|a_{D1T,n}^{(i)}(t)|^2] \equiv 2p_{TT}^{(i)}(t)+p_{DT}^{(i)}(t),
\end{equation}
respectively. 

For the triplets we solve eq S2 with the initial condition $b_{DD,n}(t=0)=\delta_{n,0}$,  
$b_{D0T,n}(t=0)=0$ and $b_{D1T,n}(t=0)=0 \; \forall n$, yielding the probabilities of finding 
both electrons in the dot $p_{DD}^{(tr)}(t)$, one electron in the dot and the other in the trap $p_{DT}^{(tr)}(t)$ and trap occupancy $n_{T}^{(tr)}(t)$ as

\begin{equation}
p_{DD}^{(tr)}(t)=\sum_{n=0}^{\infty}| b_{DD,n}(t)|^2,
\end{equation}

\begin{equation}
p_{DT}^{(tr)}(t)=\sum_{n=0}^{\infty}[| b_{D0T,n}(t)|^2+| b_{D1T,n}(t)|^2],
\end{equation}
and

\begin{equation}
n_{T}^{(tr)}(t)\equiv \langle \psi_{Tz}(t)|\sum_{\sigma}\hat{n}_{T \sigma}|\psi_{Tz}(t) \rangle =\sum_{n=0}^{\infty}(|b_{D0T,n}(t)|^2+|b_{D1T,n}(t)|^2)
\equiv p_{DT}^{(tr)}(t),
\end{equation}
respectively, for any of the initial triplet states. The correct normalization of the wavefunctions eqs \ref{psiSinglets-t} and \ref{psiTriplets-t}
ensures conservation of the number of electrons at all times for any of the initial states.

In a photoexcitation experiment, the probability of exciting an electron to the conduction band depends on the energy of the incident radiation and on the matrix elements of the dipole moment operator between the initial and final electronic wavefunctions. Therefore, not all the QD states will be populated with the same probability, in general, this depending on the material system under consideration.
We will present below results for the different initial states, all showing similar characteristics and yielding similar values of the trap occupancy. Then, we assume here that the six possible initial states are equally probable since this assumption does not change the conclusions of our work. The final result for the trap occupancy  in the one-step process, $n_{T1}(t)$, is calculated as

\begin{equation}
n_{T1}(t)=\frac{n_{T}^{(0)}(t)+n_{T}^{(1)}(t)+n_{T}^{(2)}(t)+3 n_{T}^{(tr)}(t)}{6},
\end{equation}
with the same expression for the probabilities. 
Consideration of the three-step process is more involved since the first electron can be found in three possible intermediate states,
namely, in the trap with probability $p_{1}$, or in the state $\epsilon_0$ of the dot with probability $p_{0}$ or in the state $\epsilon_1$ 
 with probability $1-p_{0}-p_1$. In the third step the second electron, with up or down spin, can be put in the states $\epsilon_0$ or $\epsilon_1$ of the
dot at $t=0$ and can find the first one in any of three mentioned intermediate states. Consequently, it is necessary to run calculations 
of eq S1 for five different intermediate initial conditions, of eq S2 for three initial intermediate conditions and weight the results
with probabilities $p_0$, $p_1$ or $1-p_0-p_1$, as appropriate. We will see below that the results of one-step and three-steps processes are very similar,
like for case of a single-level QD. For this reason and taking into account that for a system of more than two electrons the analysis of many-step processes 
will be a much more complex and longer task as it would require many more runs of the systems of differential equations, 
we will only consider one-step excitation of three or four electrons to the QD in the following.

\subsection{ Three and four electrons in a two-level QD connected to the trap: theory}

The possible initial configurations of three electrons in a two single-particle levels QD are the ones with z-component of the total spin $S_z=\pm \frac{1}{2}$ (Figure \ref{esq4} b)).
 In the representation of eq \ref{basis-3} the basis states with $S_z= \frac{1}{2}$ are

\begin{eqnarray}
|\varphi_{DD0,n}\rangle &= &|1 1; 1 0; 0 0\rangle \otimes|n \rangle,  \nonumber \\
|\varphi_{DD1,n}\rangle &= &|1 0; 1 1; 0 0\rangle \otimes|n \rangle,  \nonumber \\
|\varphi_{D00T,n}\rangle &= &|1 1; 0 0; 1 0\rangle \otimes|n \rangle,  \nonumber \\
|\varphi_{D11T,n}\rangle &= &|0 0; 1 1; 1 0\rangle \otimes|n \rangle,  \nonumber \\
|\varphi_{a,n}\rangle &= &|1 0; 0 1; 1 0\rangle \otimes|n \rangle,  \nonumber \\
|\varphi_{b,n}\rangle &= &|0 1; 1 0; 1 0\rangle \otimes|n \rangle,  \nonumber \\
|\varphi_{c,n}\rangle &= &|1 0; 1 0; 0 1\rangle \otimes|n \rangle,  \nonumber \\
|\varphi_{D0TT,n}\rangle &= &|1 0; 0 0; 1 1\rangle \otimes|n \rangle,  \nonumber \\
|\varphi_{D1TT,n}\rangle &= &|0 0; 1 0; 1 1\rangle \otimes|n \rangle.  \nonumber \\
\label{basis-3e}
\end{eqnarray}

In eq \ref{basis-3e} the first two lines represent the states of three electrons in the dot 
, the following five lines describe two electrons 
in the dot and one in the trap and the two final lines one electron in the dot and two in the trap. 
These nine states  are coupled by $H_{hop}$ and should be combined in the time dependent wavefunction as 

\begin{eqnarray}
|\psi(t) \rangle = \sum_{n=0}^{\infty}&[&a_{DD0,n}(t)|\varphi_{DD0,n}\rangle +a_{DD1,n}(t)|\varphi_{DD1,n}\rangle \nonumber \\
                                     & &+a_{D00T,n}(t)|\varphi_{D00T,n}\rangle+a_{D11T,n}(t)|\varphi_{D11T,n}\rangle \nonumber \\
                                     & &+a_{a,n}(t)|\varphi_{a,n}\rangle+a_{b,n}(t)|\varphi_{b,n}\rangle+ a_{c,n}(t)|\varphi_{c,n}\rangle \nonumber \\
                                     & &+a_{D0TT,n}(t)|\varphi_{D0TT,n}\rangle + a_{D1TT,n}(t)|\varphi_{D1TT,n}\rangle],
\label{psi3e-t}
\end{eqnarray}

The system of the nine differential equations we have to solve is given by eq S3 in the Supporting Information.
We have two possible initial states, namely $|\varphi_{DD0,n=0}\rangle$ or $|\varphi_{DD1,n=0}\rangle$ (see Figure \ref{esq4} b). After solving eq S3 
for each of the two possible initial states $i=0, 1$ with the
corresponding initial condition, we calculate 
the probabilities of finding two electrons in the trap and the other electron in the dot, $p_{TT}^{(i)}(t)$, 
all the three electrons in the dot, $p_{DD}^{(i)}(t)$, and one electron in the trap and
the other two electrons in the dot, $p_{DT}^{(i)}(t)$, and the trap occupancy $n_{T}^{(i)}(t)$ as

\begin{equation}
p_{TT}^{(i)}(t)=\sum_{n=0}^{\infty}[|a_{D0TT,n}^{(i)}(t)|^2+|a_{D1TT,n}^{(i)}(t)|^2],
\end{equation}

\begin{equation}
p_{DD}^{(i)}(t)=\sum_{n=0}^{\infty}[|a_{DD0,n}^{(i)}(t)|^2+|a_{DD1,n}^{(i)}(t)|^2],
\end{equation}

\begin{equation}
p_{DT}^{(i)}(t)=\sum_{n=0}^{\infty}[|a_{D00T,n}^{(i)}(t)|^2+|a_{D11T,n}^{(i)}(t)|^2+|a_{a,n}^{(i)}(t)|^2+|a_{b,n}^{(i)}(t)|^2+|a_{c,n}^{(i)}(t)|^2],
\end{equation}
and

\begin{equation}
n_{T}^{(i)}(t)=2 p_{TT}^{(i)}(t)+p_{DT}^{(i)}(t),
\end{equation}
respectively.  Assuming that both initial states are equally probable, the trap occupancy in the one-step process is

\begin{equation}
n_{T}(t)=\frac{n_{T}^{(0)}(t)+n_{T}^{(1)}(t)}{2},
\end{equation}
with the same expression for the probabilities.

Now we consider four electrons. The only possible initial state in a one-step process is the one with two electrons of different spins in $\epsilon_0$ and two 
electrons of different spins in $\epsilon_1$ (Figure \ref{esq4} c)). This state couples to the following states

\begin{eqnarray}
|\varphi_{DD0,n}\rangle &= &|1 1; 1 1; 0 0\rangle \otimes|n \rangle,  \nonumber \\
|\varphi_{D0T,n}\rangle &= &\frac{1}{\sqrt 2}[|1 0; 1 1; 0 1\rangle -|0 1; 1 1; 1 0\rangle ]\otimes|n \rangle,  \nonumber \\
|\varphi_{D1T,n}\rangle &= &\frac{1}{\sqrt 2}[|1 1; 1 0; 0 1\rangle -|1 1; 0 1; 1 0\rangle ]\otimes|n \rangle,  \nonumber \\
|\varphi_{D00TT,n}\rangle &= &|1 1; 0 0; 1 1\rangle \otimes|n \rangle,  \nonumber \\
|\varphi_{D11TT,n}\rangle &= &|0 0; 1 1; 1 1\rangle \otimes|n \rangle,  \nonumber \\
|\varphi_{D01TT,n}\rangle &= &\frac{1}{\sqrt 2}[|1 0; 0 1; 1 1\rangle -|0 1; 1 0; 1 1\rangle ]\otimes|n \rangle.  \nonumber \\
\label{basis-4e}
\end{eqnarray}

The system of differential equations is given by eq S4 in the Supporting Information. Proceeding as previously, we calculate 
the probabilities of finding two electrons in the trap and two electrons in the dot, $p_{TT}(t)$, 
all the four electrons in the dot, $p_{DD}(t)$, and one electron in the trap and
three electrons in the dot, $p_{DT}(t)$, and the trap occupancy $n_{T}(t)$ as

\begin{equation}
p_{TT}(t)=\sum_{n=0}^{\infty}[| a_{D00TT,n}(t)|^2+| a_{D11TT,n}(t)|^2+| a_{D01TT,n}(t)|^2],
\end{equation}

\begin{equation}
p_{DD}(t)=\sum_{n=0}^{\infty}| a_{DD0,n}(t)|^2,
\end{equation}

\begin{equation}
p_{DT}(t)=\sum_{n=0}^{\infty}[| a_{D0T,n}(t)|^2+| a_{D1T,n}(t)|^2],
\end{equation}
and

\begin{equation}
n_{T}(t)=2 p_{TT}(t)+p_{DT}(t),
\end{equation}
respectively.


\section{Results and Discussion}


In the calculations, we take $\omega_0$ as unit of energy and $\epsilon_0=0$. 
The experimental photoluminesce lineshape of ZnO can be fitted with two phonons of energies 32 meV and 40 meV \cite{Janotti_PRL}, so we consider one phonon with the average energy $\omega_0= 35 $meV  ($\omega_0=53 $ ps$^{-1}$). 
Also the values of $\frac{\lambda}{\omega_0}$ for the fitting are nearly the same for the two phonons and equal to 3.9, but in this paper we will use  $\frac{\lambda}{\omega_0}$=3 instead. This is based on the results of our previous work \cite{PRB_Monreal}, where we analyzed the case of a single electron in the system, 
with  $0 \leq \frac{\lambda}{\omega_0} \leq 6 $ and varying the inter-level energy spacing of the QD, that showed a maximum of trap occupancy for  $3 \lesssim \frac{\lambda}{\omega_0} \lesssim 4$  in all the cases. Since higher values of  $\lambda$ requires including a larger number of phonons in the calculations, which in turn increases the computer time by a large factor,  we use  $\frac{\lambda}{\omega_0}$=3 and are confident that the conclusions of the present work will not change for  $\frac{\lambda}{\omega_0}$=4. In the same work we also  showed
that the main results do not depend essentially on the values of $\tilde \epsilon_T$ and $V_{DT}$,  
and use $\tilde \epsilon_T=-10 \omega_{0}$ and $V_{DT}=\omega_0$ as in \cite{PRB_Monreal}. 
The value of $U$ will depend on the atomistic nature of the trap and its surroundings so we perform calculations for $0<U<\infty$. 
Nevertheless, a very rough estimate of $U$ is as follows. 
 Assuming a hydrogenic wavefunction for an energy
level of $\tilde \epsilon_T=-10 \omega_{0}$ we find $U= 78 \omega_0$. However this value has to be screened and very different 
values of the dielectric constant $\epsilon_s$ can be found in the literature. For bulk ZnO $\epsilon_{s}=8-10$ \cite{Hirashi,Franco} 
but it can range from 5 to 50 for nanoparticles \cite{Franco, Ahmad, Kaur}, depending on their size.
Hence screened values of $U$ could range between $2 \omega_0$ and $15 \omega_0$. 
The time-dependent equations are solved using a standard Runge-Kutta method. The value of the time step depends on the parameters and the calculations are checked by ensuring the correct normalization of all the wavefunctions with an accuracy better than 1$\%$, $ \forall t$. The number of phonons that should be included in the calculation obviously depends on $\lambda$ and we need to include typically 3-4$n_b(t)$. This indicates that a large number of phonons are virtually involved in the process.
The calculations run up to a maximum time $\omega_0 t_{max}=200$ which is enough to have converged results. 

\subsection{A single-level QD connected to the trap: results} 

\begin{figure}
\centering
\includegraphics[width=100mm]{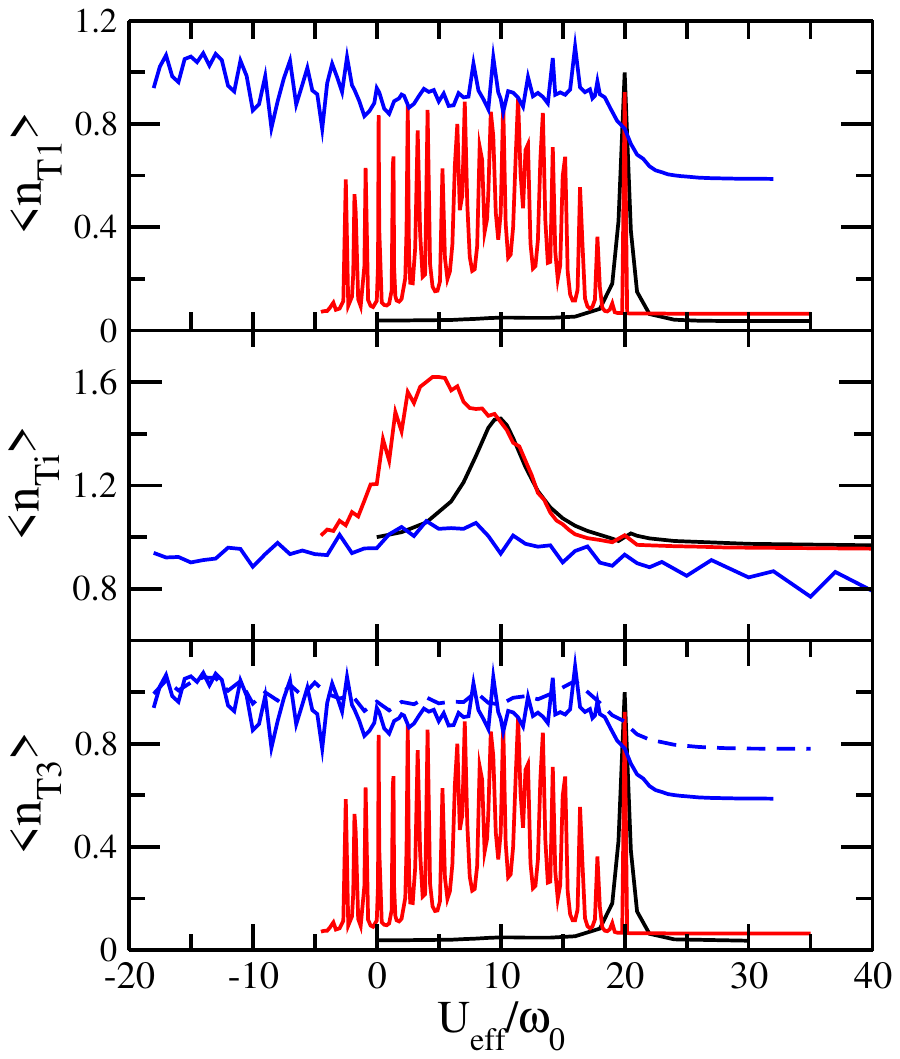}
\caption{The time averaged trap occupancies for a system of two electrons in a single-level QD connected to the trap as a function of $U_{eff}$, 
for the one-step process $<n_{T1}>$  (upper panel) and the initial intermediate singlet state in the three-step process having one electron in the dot and the other in the trap  
$<n_{Ti}>$ (middle panel).
The lower panel compares the results for the one-step  (continuous lines) and the three-step (eq 16, dashed lines) processes. $V_{DT}=\omega_0$.
Black lines: $\lambda=0$, red lines:
$\lambda=1.5\omega_0$ and blue lines: $\lambda=3\omega_0$. For $\lambda=0$ and $\lambda=1.5\omega_0$ 
the results of the one-step and three-step processes are indistinguishable on the scale of the figure
.} 
\label{fig1}
\end{figure}

The upper panel of Figure \ref{fig1} shows the time-averaged trap occupancy as a function of $U_{eff}$ ($U_{eff}=U-2\frac{\lambda^2}{\omega_0}$) 
for the case of two electrons in a single-level QD, one-step process, $V_{DT}=\omega_0$ \cite{PRB_Monreal} and three values of $\lambda$. 
In the absence of electron-phonon coupling ($\lambda=0$) the occupancy is smaller than 0.5$\%$ for all values of $U_{eff}$ except for $U_{eff} \simeq 20\omega_0$, 
for which the energy of the two electrons in the trap $2\tilde \epsilon_T+U_{eff} \simeq 0$ equals the energy of the two electrons in the dot. 
In this case our calculation shows that both electrons jump together back and forth from dot to trap periodically, so the trap occupancy oscillates in time from 0 to 2. 
For $\lambda=1.5 \omega_0$ we still see the same
peak and also other very narrow ones at values of $\tilde \epsilon_T+U_{eff} \simeq m\omega_0$, with $m$ an integer (see eq 5). These peaks are superimposed on a continuum with a
maximum at $U_{eff} \simeq 10\omega_0$ of 35 $\%$, and $<n_{T1}>$ is larger than for $\lambda=0$. Further increase of $\lambda$ moves this maximun to lower values of $U_{eff}$ while increasing in intensity 
and for $\lambda=3 \omega_0$ the trap occupancy is about 1 for negative values of $U_{eff}$ and about 0.9 for  $0 \lesssim U_{eff} \lesssim 20\omega_0$. 
For larger values of $U_{eff}$ $<n_{T1}>$  decreases step-like, approaching quickly
its $U\rightarrow \infty$ limit where one electron at most can occupy the trap state at any time and, consequently, $p_{TT}(t) \rightarrow 0$. 
Actually, in the $U\rightarrow \infty$ limit, eq \ref{equ2-t} reduce to the ones describing a single electron 
in the system \cite{PRB_Monreal} with a (renormalized) hopping parameter $\sqrt2 V_{DT}$ \cite{Ashcroft}.
The middle panel of Figure \ref{fig1}  shows the time-averaged trap occupancy for the initial intermediate singlet state of the three-step process having one electron in the dot and the other in the trap, $<n_{Ti}>$ in eq 16, and the same values of $\lambda$. ($n_{b1}=0$ for 
$\lambda=0$ and $1.5 \omega_0$ and $n_{b1}=12$ for $\lambda=3 \omega_0$).
In this case the main peak occurs at $\tilde \epsilon_T+U_{eff} \simeq 0$, that is when the energy of 
the second electron in the trap $\tilde \epsilon_T+U_{eff}$ equals its energy in the dot. Finally, the lower panel compares the results of the one-step and three-step process, 
as calculated by eq 16. For $\lambda=3 \omega_0$ $p_{1}=0.6$ and $\langle n_{T1} \rangle$ and $\langle n_{T3} \rangle$ only differ
appreciably for large values of $U_{eff}$. For $\lambda=1.5 \omega_0$ $p_{1}=0.04$ and for $\lambda=0$ $p_{1}=0.02$ so the occupancies for both kind of process are 
nearly the same for any $U_{eff}$ and indistinguishable on the scale of the Figure.
At this point we should mention that we have performed calculations for $\tilde \epsilon_T=-10.5 \omega_{0}$ and $\lambda=3 \omega_0$ confirming the insensibility of the results to the change in $\tilde \epsilon_T$ that we found previously for the single electron case. 
Increasing $\lambda$ to $\lambda=4 \omega_0$ keeps similar values of $\langle n_{T1} \rangle$ for $U_{eff} \ge 0$
and decreases to $\langle n_{T1} \rangle  \simeq 0.85$ for $U_{eff} \le 0$. These results are presented in Figure S1 of the Supporting Information. Then, similar to the single electron case analyzed in \cite{PRB_Monreal}, the maximum of the trap
occupancy occurs for $3\omega_0 \leq \lambda \leq 4 \omega_0$.

\begin{figure}
\centering
\includegraphics[width=100mm]{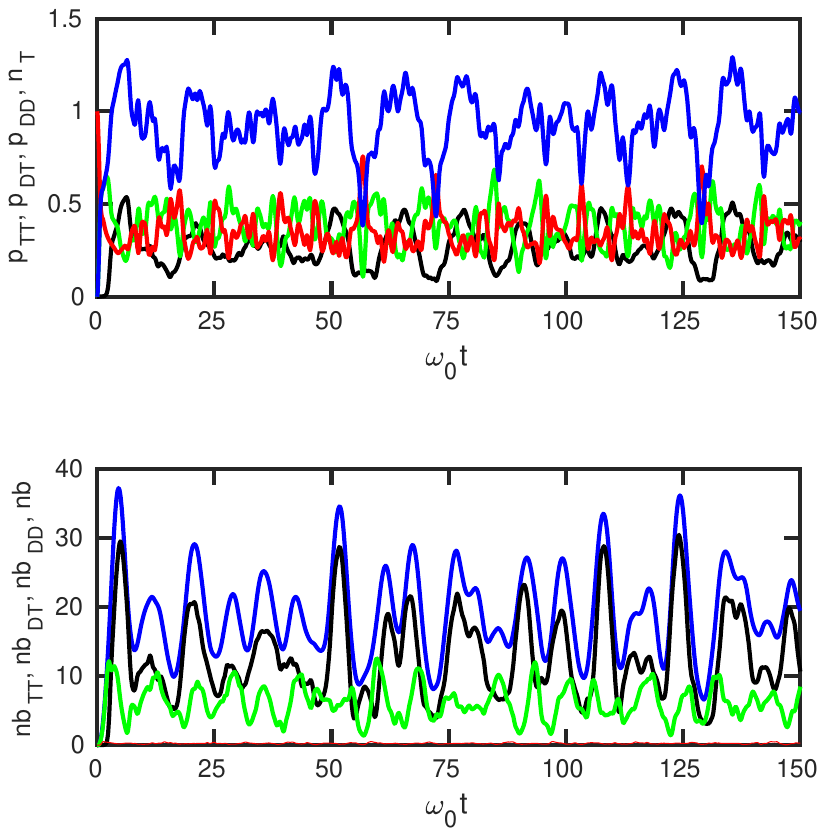}
\caption{Time dependent probabilities and trap occupancy (upper panel) and their associated number of phonons (lower panel) 
for a system of two electrons in a single level QD, one-step process, $\lambda= 3 \omega_0$ and $U_{eff}= 11 \omega_0$, $V_{DT}=\omega_0$. Black lines: $p_{TT}$, red lines: $p_{DD}$, green lines: $p_{DT}$ and
blue lines $n_{T1}$. The number for phonons associated to $p_{DD}$ is so small that cannot be seen on the scale of the Figure
.} 
\label{fig2}
\end{figure}

An important conclusion of the present calculation is that the trap occupancy of the two-electron system is much higher than in the single electron case for any value of the electron-phonon coupling $\lambda$ independently of $U_{eff}$, as long as $U_{eff} \lesssim -2 \tilde \epsilon_T$. 
In particular for $\lambda=1.5 \omega_0$ the trap occupancy of a single electron is $\simeq 0.04$, much smaller than the values shown in Figure \ref{fig1}, 
and for $\lambda \simeq 3 \omega_0$, where we found the maximum of the single electron trap occupancy to be $\langle n_{T} \rangle  \simeq 0.6$, 
we find $\langle n_{T} \rangle \simeq 1 $ for the two-electron system. This means that the trap photoluminescence 
will be completely quenched by adding one more electron to the QD. 
The reason for this increase can be inferred from Figure \ref{fig2}, where we show the time-dependent probabilities, trap occupancy and number of phonons for the one-step process, 
$\lambda= 3 \omega_0$ and $U_{eff}= 11 \omega_0$. The probability of having the two electrons in the trap, 
$p_{TT}(t)$ undergoes fast oscillations in time around its average value of 0.27,
while $p_{DT}$ and $p_{DD}$ oscillate around average values of 0.38 and 0.35, respectively. These oscillations are basically the same Rabi oscillations that we found for the case of a single electron in the system, reflecting coherence between electronic states \cite{PRB_Monreal}.
We see that that within times $\omega_{0}t \simeq 10-20$ ($t \simeq 0.2-0.4$ ps for $\omega_0= 53$ps$^{-1}$), the three possible states of the two-electron system, eq \ref{basis-2}, 
are nearly equally probable and, 
consequently, $\langle n_{T1}\rangle = \langle 2p_{TT}+p_{DT} \rangle \simeq 1 $. Moreover the number of phonons in the state with the two electrons in the trap
is the largest at any time. This is seen for many values of $U_{eff}$, as well. Finally, notice the plateaux in $n_{T1}(t)$ where it is almost constant during time intervals 
$\omega_{0} t \simeq 25-30$.


\subsection{A two-level QD connected to the trap: results}

We now present our results for a two-level QD housing 2 to 4 electrons. The chosen values of $\epsilon_1-\epsilon_0$ are in the same range as these
of ZnO nanospheres of radii between 2 and 6 nm: $0.5 \omega_0 \lesssim \epsilon_1-\epsilon_0 \lesssim 2 \omega_0$.
To make contact with our previous work \cite{PRB_Monreal}, the QD is connected to the trap by the hopping parameter $V_{DT}=\frac{1}{\sqrt 2}\omega_0$, corresponding to
$V_0=\omega_0$ in \cite{PRB_Monreal}, and concentrate on the case $\lambda=3 \omega_0$
because we showed that the electron-phonon interaction is more efficient for $3 \omega_0 \lesssim \lambda \lesssim 4 \omega_0$. We also run calculations for all values of $U$. 
We should mention that the time-dependent probabilities behave very much like the ones shown in Figure \ref{fig2} and therefore their time-averaged values 
are also set on a subpicoseconds timescale.
From now on, and in order to simplify the notation,
time averaged values will
be denoted by the corresponding symbols without angle brackets (example: $p_{TT} \equiv <p_{TT}(t)>$) unless otherwise indicated.

\begin{figure}
\centering
\includegraphics[width=100mm]{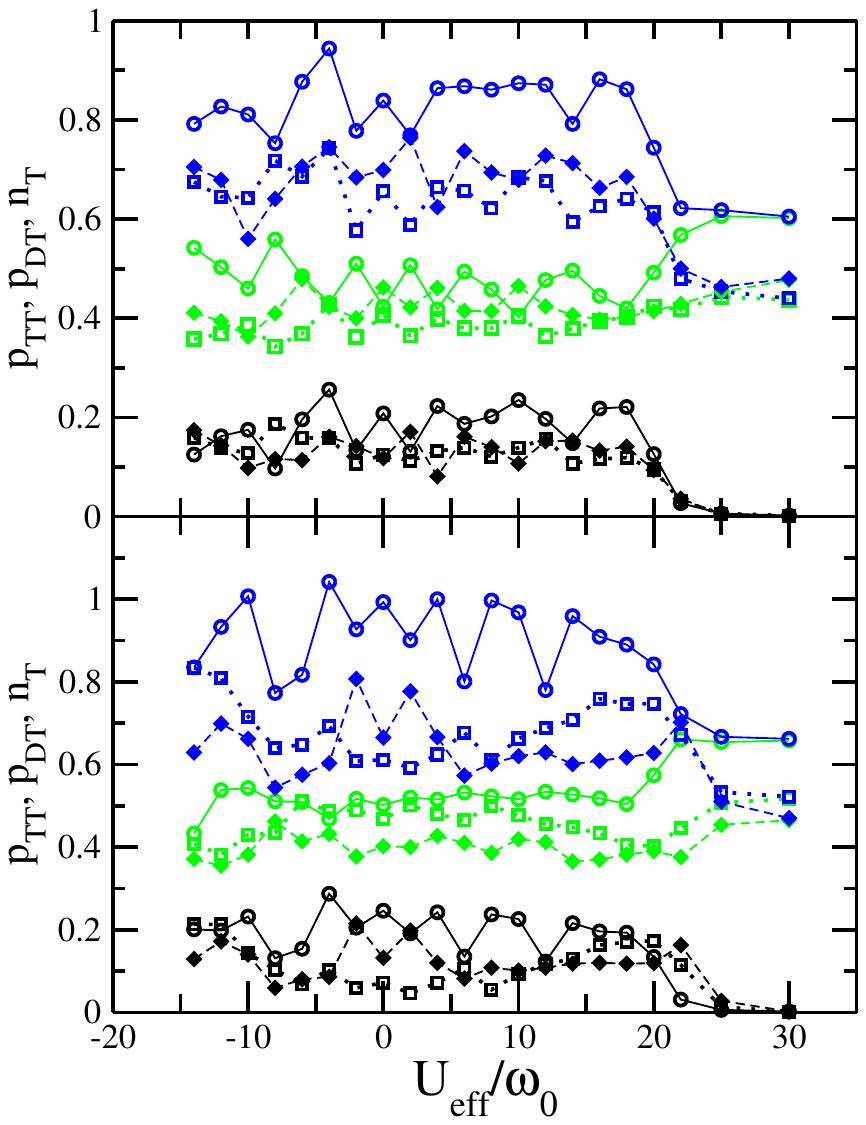}
\caption{The time averaged probabilities $p_{TT}$ (black symbols), $p_{DT}$ (green symbols),
and trap occupancies $n_{T}$ (blue symbols) as a function of $U_{eff}$ for two electrons in a two-level QD, $\epsilon_1=0.5 \omega_0$ (upper panel)
and $\epsilon_1=2 \omega_0$ (lower panel), $\lambda=3 \omega_0$, $V_{DT}=\frac{1}{\sqrt 2}\omega_0$. Calculations are for the three initially possible singlet states 
 $|S_0\rangle $ (open dots with full line), $|S_1\rangle $ (open squares with dotted line) and $|S_2\rangle $ (full diamonds with dashed line)
.} 
\label{fig3}
\end{figure}

 Starting with a system of two electrons, Figure \ref{fig3} illustrates the dependence of the time-averaged probabilities and trap occupancies on the initial singlet state
 for two values of $\epsilon_1$ differing by a factor of four. While $p_{DT}$ is, in general, rather constant with $U_{eff}$ and 
 do not depend strongly on the initial state, $p_{TT}$ and $n_T$, present large variations with $U_{eff}$ and depend more strongly on the initial state.
 As an example, $n_T$ can change from 0.6 for $|S_1\rangle $ to 0.8-1.0 for $|S_0 \rangle $ at some values of $U_{eff}$.
 The largest values of $n_T$ are always obtained for $|S_0\rangle $, that is when the two electrons are initially in the ground state of the QD but 
 there is no systematic behavior with the other two initial states: $n_T$ for $|S_1\rangle $ can be larger than for $|S_2\rangle $ at some values 
 of $U_{eff}$ but smaller at others. As a reference, we quote the values of $n_T$ for any of the initial triplets: $n_{T}^{(tr)}$= 0.742 for $\epsilon_1=0.5 \omega_0$ and 
 $n_{T}^{(tr)}$= 0.627 for $\epsilon_1=2 \omega_0$, similar to the singlet values $n_{T}^{(1)}$ or $n_{T}^{(2)}$ but smaller than $n_{T}^{(0)}$.
 The values of $U_{eff}$ making the step-like transition to the $U \rightarrow \infty$ limit increase with 
 the energy of the initial state as $U_{eff} \simeq -2 \tilde \epsilon_T+E^{(i)}$, where 
 $E^{(i)}$ is the energy of the initial state $|S_0\rangle $, $|S_1\rangle $ or $|S_2\rangle $.

\begin{figure}
\centering
\includegraphics[width=100mm]{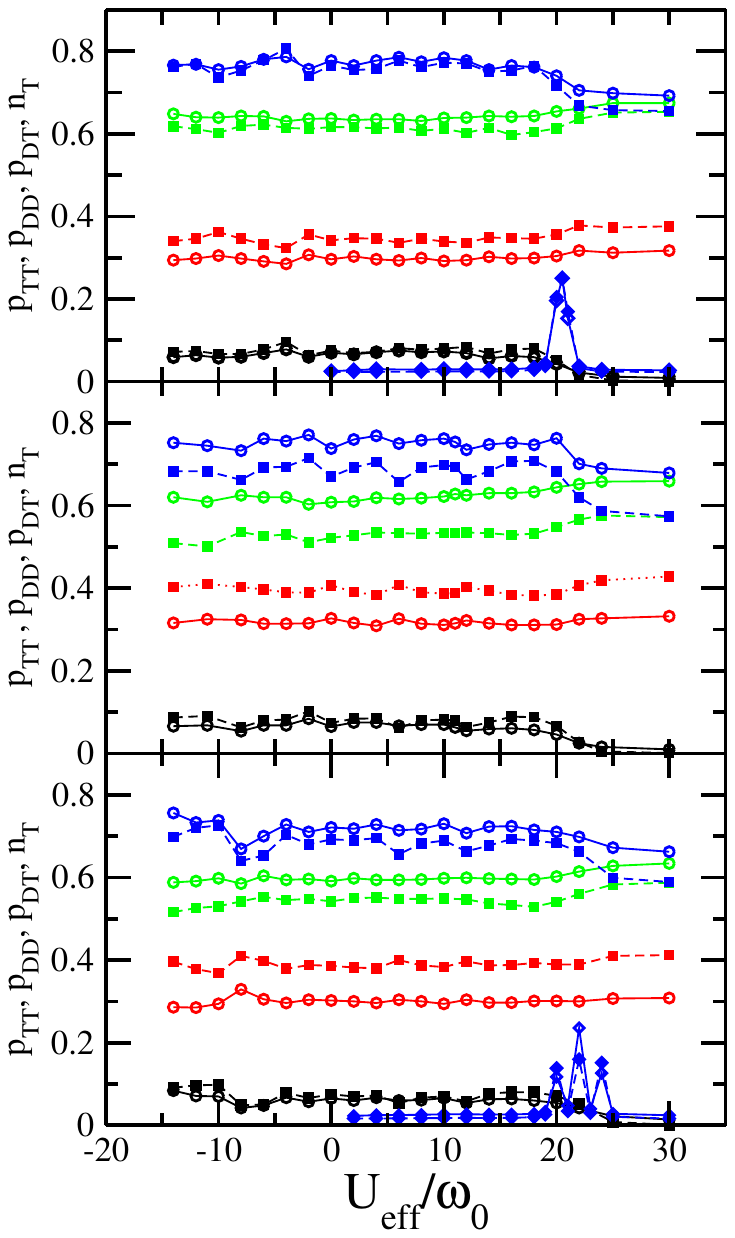}
\caption{The time averaged probabilities $p_{TT}$ (black symbols), $p_{DD}$ (red symbols), $p_{DT}$ (green symbols),
and trap occupancies $n_{T}$ (blue symbols), as a function of $U_{eff}$ for two electrons in a two-level QD and 
$\epsilon_1=0.5 \omega_0$ (upper panel),  $\epsilon_1=\omega_0$ (middle panel)
and $\epsilon_1=2 \omega_0$ (lower panel) and for the one-step (full squares with dashed line) and three-step (open dots with continuous line) processes.
$\lambda=3 \omega_0$, $V_{DT}=\frac{1}{\sqrt 2}\omega_0$.
Blue diamonds show the trap occupancies for $\lambda=0$
.} 
\label{fig4}
\end{figure}

Figure \ref{fig4} shows time averaged probabilities $p_{TT}$, $p_{DD}$, and $p_{DT}$ and the trap occupancies $n_{T}$ for three values of $\epsilon_1$, comparing  the 
one-step and the three-step processes. We can observe many characteristics common to the case of two electrons in a single-level QD presented in Figure \ref{fig1},
that are independent of $\epsilon_1$. Probabilities
and occupancy depend weakly on $U_{eff}$ for $U_{eff} \lesssim -2 \tilde \epsilon_T$ but decrease for larger values making a quick transition to 
the $U \rightarrow \infty$ limit. No significant differences between one-step and three-step processes are 
observed although the later process always gives somewhat larger values. 
The Figure also displays $n_T$ for $\lambda=0$, which only reaches appreciable values for values of $U_{eff}$ such as
the energy of the two electrons in the trap, $U_{eff}+2 \tilde \epsilon_T$, equals the energy of two electrons in the dot, $2\epsilon_0$, $\epsilon_0+\epsilon_1$
or $2\epsilon_1$. However, $n_T \approx 0.7-0.8$ for $\lambda=3 \omega_0$ independent of $U_{eff}$, this showing the importance of strong electron-phonon interaction.
Although the values of $p_{TT}$ shown in Figure \ref{fig3} for the initial singlet states do not differ much from the ones found for the single-level QD,
the one-step values of $p_{TT}$ in Figure \ref{fig4} are typically
a factor of two smaller. This is due to the dilution caused by the triplet states, for which $p_{TT}=0$, in eq  31. 
An important conclusion of the present calculation is, again, that adding one more electron to the system increases the occupancy of the trap state by a large factor. Specifically, for $\epsilon_1= 0.5 \omega_0$ the occupancy of the trap increases from 0.39 for one electron to 0.7-0.8 for two electrons in the QD,
while for $\epsilon_1= 2 \omega_0$ it increases from 0.48 to 0.7 when increasing from one to two electrons.

We now present our results for a system of three and four electrons in a two-level QD.

\begin{figure}
\centering
\includegraphics[width=100mm]{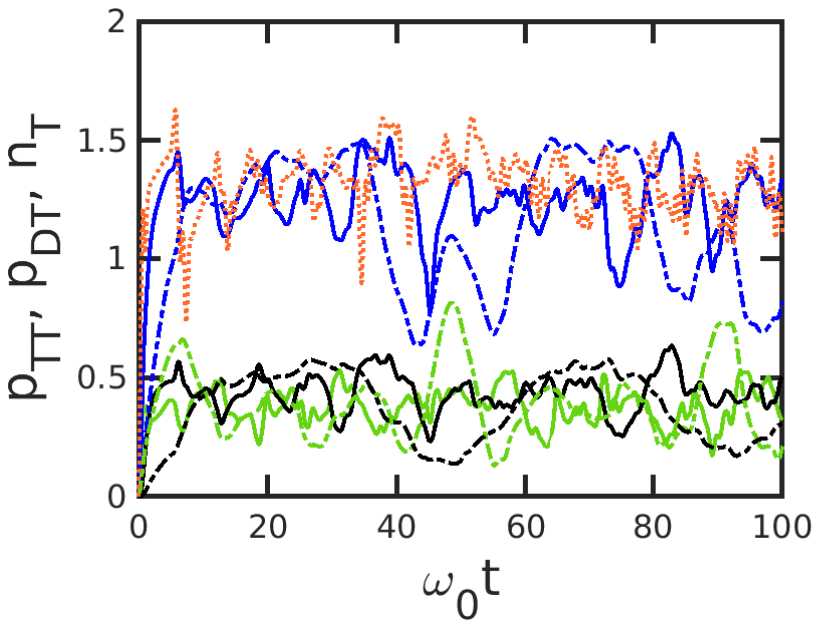}
\caption{ Time-dependent probabilities $p_{TT}$ (black lines), $p_{DT}$ (green lines) and trap occupancies $n_{T}$ (blue lines), 
for a system of four electrons in a two-level QD with $U_{eff}=-12 \omega_0$, $\epsilon_1=2 \omega_0$, $\lambda=3 \omega_0$ and $V_{DT}=\frac{\omega_0}{\sqrt{2}}$ 
(continuous lines)
and $V_{DT}=\frac{\omega_0}{2 \sqrt{2}}$ (dashed lines). The orange dotted line is the trap occupancy for $V_{DT}=\frac{2 \omega_0}{\sqrt{2}}$
.} 
\label{fig5}
\end{figure}

Figure \ref{fig5} displays the time-dependent probabilities and trap occupancy of a system of four electrons in a QD connected to the trap by two
values of the hopping parameter $V_{DT}=\frac{\omega_0}{\sqrt{2}}$ (continuous lines)
and $V_{DT}=\frac{\omega_0}{2 \sqrt{2}}$ (dashed lines). The trap occupancy for $V_{DT}=\frac{2 \omega_0}{\sqrt{2}}$ is also shown by the orange dotted line.
Notice how the probabilities of having two electrons in the trap (and the other two in the dot) and one electron in the trap (and the other three in the dot) 
oscillate around their mean values of ca. 0.4, giving trap occupancies larger than 1.2, 
and these mean values are obtained in timescales of  0.4-0.8 ps, similar to the case of two electrons in a single level QD shown in Figure \ref{fig2} .
This behavior, with similar values, is seen even if we decrease or increase the the hopping parameter by a factor of two, 
so we expect it to be quite independent of $V_{DT}$ like in in the single-electron case \cite{PRB_Monreal}. Of course, this only holds for values of $V_{DT}$ 
on the order or larger than $\epsilon_1$. Note also how the trap occupancy stays 
almost constant in time for $V_{DT}=\frac{2 \omega_0}{\sqrt{2}}$.

\begin{figure}
\centering
\includegraphics[width=100mm]{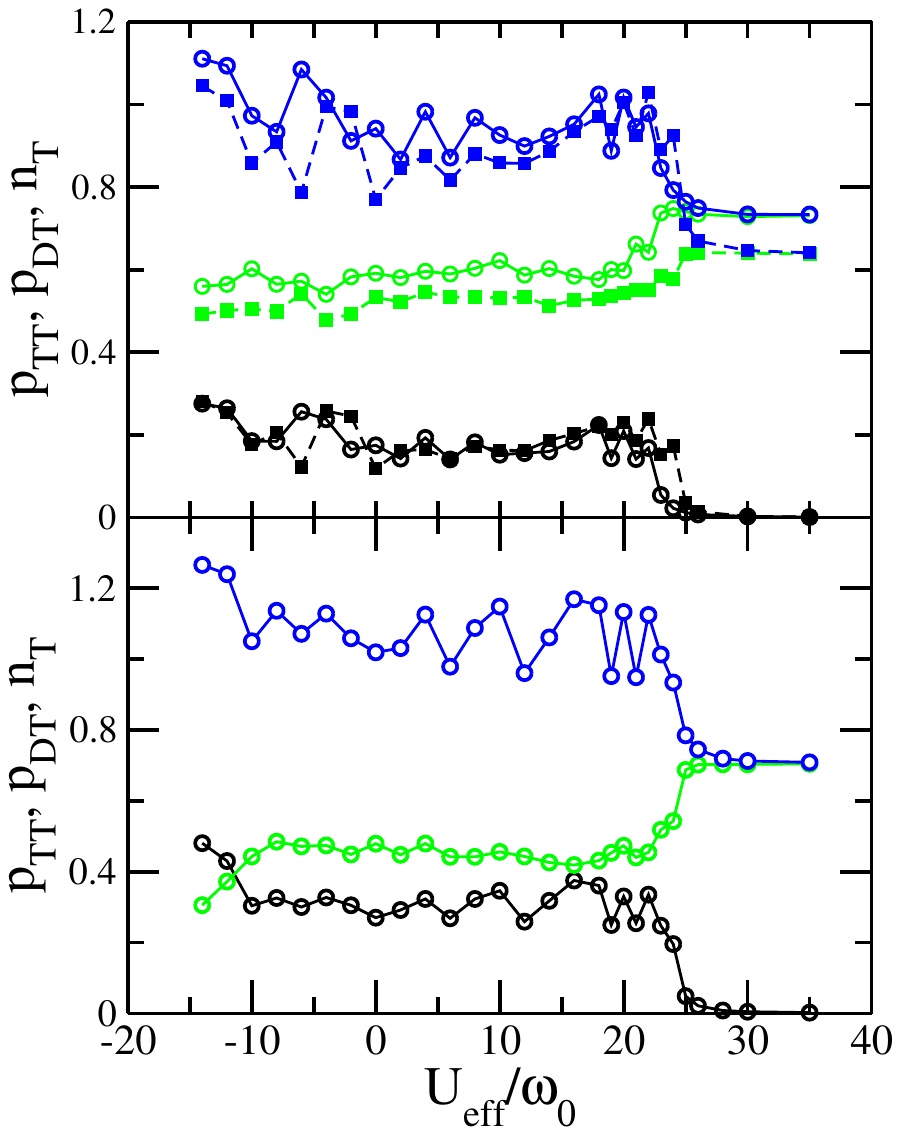}
\caption{The time averaged probabilities $p_{TT}$ (black symbols), $p_{DT}$ (green symbols) and trap occupancies $n_{T}$ (blue symbols), 
as a function of $U_{eff}$ for a system of three electrons (upper panel) 
and four electrons (lower panel) in a two-level QD with $\epsilon_1=2 \omega_0$, $\lambda=3 \omega_0$, $V_{DT}=\frac{1}{\sqrt 2}\omega_0$. 
The upper panel shows results for the two possible initial states 
 $i=0 $ (open dots with full line) and $i=1 $ (open squares with dotted line) of the three electron system 
.} 
\label{fig6}
\end{figure}

Figure \ref{fig6} shows our results for a system of three and four electrons. Also shown are results for the two possible initial states in the three electron system. 
The probabilities of finding all the electrons in the dot $p_{DD}$ are not shown for clarity, since they take similar values as $p_{TT}$, but can be readily obtained from the figure 
since $p_{DD}=1-p_{TT}-p_{DT}$.
As in the case of the two-electron system, one can notice the step-like decrease in $p_{TT}$ and $n_T$ 
that in the case of three or four electrons is at $U_{eff} \simeq 24 \omega_0$.
This is the value of $U_{eff}$ for which the maximum energy of the state with two electrons in the trap equals the maximum energy of the initial state.
Comparison of this Figure to Figure \ref{fig4} clearly shows that increasing the number of electrons that are initially in the dot increases the probability 
of having two electrons in the trap, $p_{TT}$, 
while the probability of having one electron in the trap and the rest of electrons in the dot $p_{DT}$ stays almost constant to $p_{DT} \approx 0.5$. 
Hence the trap occupancy increases by adding electrons to the system. 
Also notice that all the calculations yield the same value of $n_T$ in the $U \rightarrow \infty$ limit, independently of the number of electrons in the system, as it should
since in this limit one electron at most can occupy the trap state.

To summarize the most important results of the present work, we go back to the rough estimate of $U$ given above,
 $2\omega_0 \lesssim U \lesssim 15 \omega_0$, and get rough values of $U_{eff}$  as $-16 \omega_0 \lesssim U_{eff} \lesssim -3 \omega_0$ for $\lambda= 3 \omega_0$.
 Although the trap occupancy is in general somewhat larger for negative than for positive values of $U_{eff}$, we have seen that
the time-averaged magnitudes do not depend strongly on $U_{eff}$ below its step value. Then 
we have taken "$U_{eff}$-averaged" values of $p_{TT}$, $p_{TD}$, $n_T$ and the number of electrons in the dot, $n_D$,
as a function of the number of electrons initially in the dot $n_e$ ($1 \leq n_e \leq 4$) from Figures \ref{fig4} and \ref{fig6}. These values are listed in Table I. 
It is evident that the number of electrons in the trap increases with $n_e$ and one electron is transferred to the trap when three electrons are initially  in the QD in timescales of 
a few tenths of picoseconds. This process is much faster than radiative exciton decay occurring in timescales of $\simeq 20$ ps \cite{JPCC_116_20633_cohn} to $\simeq 10$ ns \cite{Efros_NatNano} and faster than Auger recombination processes \cite{Nanolett_Vaxemburg, ChemRev_Guyot}, 
giving way to the complete suppression of the luminescence to this state.  
At the same time electrons accumulate in the dot, forming the ultralow-density gas of conduction band electrons 
characteristic of semiconductor QDs.
This facts are made possible by an efficient electron-phonon interaction. Finally, we would like to point out that  
calculations of the photoluminescence lineshapes of several defects in GaN and ZnO yield values of the phonon frequencies between 21 meV and 47 meV with values of 
$\frac{\lambda}{\omega_0}$ between 3.2 and 6 \cite{Janotti_PRL, Janotti_APL}.  Simulations for Cu-doped CdSe nanocrystals give vibrational energies of 24.3 meV and 26 meV with $\frac{\lambda}{\omega_0}$ between 1.5 and 3  \cite{Nelson}. All these values are comparable to the ones used in the present paper and therefore we think that the trap-filling mechanism proposed here is applicable to a variety of quantum dots systems.


\begin{table}[h!]
  \begin{center}
    \caption{Probabilities of having two electrons in the trap, $p_{TT}$, one electron in the trap and the rest of the electrons in the dot, $p_{DT}$,
     number of electrons in the trap, $n_{T}$, and number of electrons in the dot, $n_{D}$, as a function of the initial number of electrons in the dot
    $n_{e}$, for the two-level QD, $\epsilon_1=2 \omega_0$, $\lambda= 3\omega_0$ and $V_{DT}=\frac{\omega_0}{\sqrt{2}}$. 
    These are "$U_{eff}$-averaged" values from Figures \ref{fig4} and \ref{fig6}.}
    \label{tab:table1}
    
    \begin{tabular} { | c | c | c | c | c |}
      \hline
      $n_e$ & $p_{TT}$ & $p_{DT}$  & $n_{T}$ & $n_{D}$\\
      \hline
      \hline
      1 & 0.00 & 0.48  & 0.48 & 0.52\\
      \hline
      2 & 0.07 & 0.55  & 0.69 & 1.31\\
      \hline
      3 & 0.20 & 0.55  & 0.95 & 2.05\\
      \hline
      4 & 0.30 & 0.50  & 1.10 & 2.90\\
      \hline
    \end{tabular}
  \end{center}
\end{table}


\section{Conclusions}
\label{sec-conclusions}

In this work we have analyzed the effects of electron-electron and electron-phonon interactions in the dynamics of a system of a few electrons that can be trapped to a localized state and detrapped to an extended band state using a simple model system. 
In spite of its simplicity the time dependent model has no analytical solution but a numerically exact one can be found at a relatively low computational cost.   
We start with the simplest possible system consisting of two electrons in a single-level QD connected to the trap. The electronic motion is quasi-periodic in time, with oscillations around a mean value that is set in a sub to picoseconds timescale. These oscillations are basically the same Rabi oscillations that we found for the case of a single electron in the system, reflecting coherence between electronic states.
Next, we consider a two-level QD housing 2 to 4 electrons and find the same kind of motion. We also follow the average number of these electrons transferred to the trap state
and  find that one electron is transferred for three electrons initially in the QD.
 At the more efficient values of the electron-phonon coupling parameter $\lambda$, these characteristics are quite independent of $U_{eff}$ and of the energy of the QD 's single-particle levels. 
 We thus conclude that strong electron-phonon interaction is an efficient mechanism that can provide the complete filling of a deep trap state on a sub to picoseconds timescale, faster than radiative exciton decay and Auger recombination processes. 
 This leads to the complete suppression of the luminescence to the deep trap state and to the accumulation 
of electrons in the QD, forming the ultralow-density gas of conduction band electrons 
characteristic of semiconductor QDs \cite{JPCL_4_3024_faucheux, JPCL_5_976_faucheaux, ACSNano_8_1065_schimpf}.

\section{Acknowledgments}
 Financial support from the Spanish
Ministry of Science and Innovation through the Mar\'ia de Maeztu Programme for Units of Excellence in R$\&$D
(CEX2018-000805-M) and the project PID2021-126964OB-I00 is acknowledged. 

\begin{suppinfo}
The Supporting Information gives the sets of coupled linear differential equations for the combination coefficients in the wavefunction expansions that we have to solve in each of the analyzed cases. Additional calculations of the time-averaged trap occupancy for other values of $\tilde \epsilon_T$ and $\lambda$ are presented in Figure S1.
\end{suppinfo}

\begin{figure}
\centering
\includegraphics{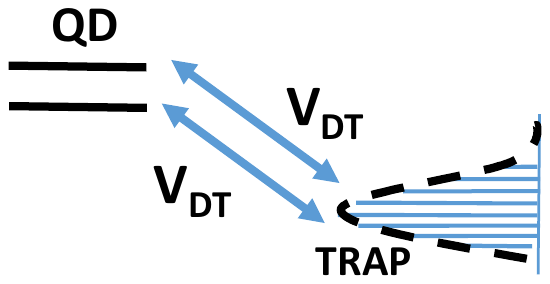}
\label{TOC_Graphic}
\end{figure}

\end{document}